\documentclass{article}
\usepackage[latin1]{inputenc}
\usepackage{amsmath}
\usepackage{amsfonts}
\usepackage{amssymb}
\usepackage{latexsym}
\usepackage{mathrsfs}
\usepackage{twistor}

\usepackage[pdftex]{hyperref}

\begin{document}

\title{\textsf{The Generalized Good Cut Equation}}
\author{\small T.M. Adamo\footnote{\texttt{adamo@maths.ox.ac.uk}} \\
\small \textit{Mathematical Institute}\\
\small \textit{University of Oxford}\\
\small \textit{24-29 St Giles}\\
\small \textit{Oxford, OX1 3LB, UK} 
\and \small E.T. Newman\footnote{\texttt{newman@pitt.edu}} \\
\small \textit{Department of Physics \& Astronomy}\\
\small \textit{University of Pittsburgh}\\
\small \textit{3941 O'Hara Street}\\
\small \textit{Pittsburgh, PA 15260, USA}}
\maketitle

\begin{abstract}
The properties of null geodesic congruences (NGCs) in Lorentzian manifolds
are a topic of considerable importance. More specifically NGCs with the
special property of being shear-free or asymptotically shear-free (as either
infinity or a horizon is approached) have received a great deal of recent
attention for a variety of reasons. \ Such congruences are most easily
studied via solutions to what has been referred to as the 'good cut
equation' or the 'generalization good cut equation'. It is the purpose of
this note to study these equations and show their relationship to each
other. \ In particular we show how they all have a four complex dimensional
manifold (known as $\mathcal{H}$-space, or in a special case as complex
Minkowski space) as a solution space.
\end{abstract}

\tableofcontents

\newpage

\section{Introduction}

Shear-free (and its generalization to asymptotically shear-free) null
geodesic congruences (NGCs) in Lorentzian space-times have played a variety
of important roles in space-time geometry. \ They first appeared in the
search for algebraically special solutions of Maxwell's equations in curved
space-times \cite{Rob61}. This was followed by the discovery of the vacuum
twist-free algebraically special metrics of Robinson and Trautman \cite{RT62}
and the extension, via the remarkable Goldberg-Sachs theorem \cite{GS62}, to
all algebraically special vacuum metrics. Penrose then showed the close
connection of shear-free congruences, via the Kerr theorem, with twistor
theory \cite{Penrose67}. More recently the asymptotically shear-free
congruences (as null infinity is approached) have been used to give physical
interpretations to the asymptotic fields in asymptotically flat space-times 
\cite{AKN09}. \ This latter case has been generalized to congruences that
become shear-free as a non-expanding horizon in a space-time is approached 
\cite{AN09}.

All the cases are governed by solutions to different versions of the same
type of partial differential equation, referred to as \textquotedblleft good
cut equations.\textquotedblright\ The good-cut equation (GCEq) or its
generalization, the generalized GCEq (G$^{2}$CEq), are second order PDEs
that live on 3-manifolds. Although it is very likely that the discussion
could be generalized to arbitrary 3-surfaces in a Lorentzian manifold (or
more specifically to arbitrary null surfaces), we confine ourselves to the
cases where the 3-surfaces are specifically Penrose's future null infinity, $%
\mathfrak{I}^{+}\mathfrak{,}$ or a vacuum non-expanding horizon, $\mathfrak{H%
}$ (c.f., \cite{Ash00,ABL02}) . For the general discussion we will refer to
both $\mathfrak{I\ }$and/or $\mathfrak{H}$ as $\mathfrak{N}.\ $In each of
these cases $\mathfrak{N\ }$has topology $S^{2}\times \mathbb{R}$,$\ $and is
foliated by null geodesics congruences whose shear and divergence both
vanish on $\mathfrak{N}$. \ Though these manifolds are real 3-surfaces in
the real space-time, we must consider their complexification, $\mathfrak{N}_{%
\mathbb{C}}$ (i.e., their analytic continuation, at least a small way, into
the complexification of the space-time). \ The coordinatization of $%
\mathfrak{N\ }$is given by Bondi-like coordinates: ($\zeta ,\overline{\zeta }
$), which label the null generators of $\mathfrak{N,\ }$are the
stereographic coordinates on the $S^{2}\ $portion of $\mathfrak{N\ }$($S^{2}$
need not be a metric sphere); while the coordinate $u\ $parametrizes the
cross-sections of $\mathfrak{N.}$ For $\mathfrak{N}_{\mathbb{C}},$the $u$ \
is allowed to take complex values close to the real, while $\overline{\zeta }%
\ $goes over to an independent variable $\widetilde{\zeta }\ $\textit{close}
to the complex conjugate of $\zeta .\ $To avoid the notational nuisance of
repeatedly saying that $\widetilde{\zeta }\approx \overline{\zeta }$,$\ $we
will simply use $\overline{\zeta }\ $to mean an independent complex variable
taking values at, or close to, the complex conjugate of $\zeta $. \ The
context should make its$\ $usage\ clear. The distinction between the$\ $GCEq
and the G$^{2}$CEq is that the former lives on a 3-surface $\mathfrak{N}$
whose $u=$constant\ cross-sections are metric spheres, while for the latter
equation the metric is arbitrary.

In Section 2, we first discuss the geometric meaning of the GCEq and its
connection to (asympotically) shearfree NGCs. \ This is followed by a review
of the known properties of solutions to the GCEq: its four
complex-dimensional solution space and its relationship to twistor theory. \
In Section 3 we show that the G$^{2}$CEq can be transformed directly to the
GCEq so that the solution spaces of both equations and properties are
equivalent. \ We demonstrate the utility of this result by applying it to
the twistor space associated with horizon shear-free NGCs at a non-expanding
horizon. \ Section 4 concludes with a discussion of possible applications of
our findings.

\section{The Good Cut Equations}

Before describing the GCEq we first discuss a notational choice. \ As
mentioned earlier, the 3-surface $\mathfrak{N\ }$is described by an $S^{2}\ $%
worth of null geodesics with the cross sections given by $u=$constant. The
metric of the two-surface cross-sections$\ $are expressed in stereographic
coordinates ($\zeta ,\overline{\zeta }$)\ so that the metric takes the
conformally flat form:%
\begin{equation}
ds^{2}=\frac{4d\zeta d\overline{\zeta }}{P^{2}(\zeta ,\overline{\zeta })},
\label{conformally flat}
\end{equation}%
with $P(\zeta ,\overline{\zeta })\ $an arbitrary smooth non-vanishing
function on the $(\zeta ,\overline{\zeta }),$ the extended complex plane
(Riemann sphere). In the special case of a metric sphere we take 
\[
P=P_{0}\equiv 1+\zeta \overline{\zeta }, 
\]%
while in general we write 
\begin{equation}
P=V(\zeta ,\overline{\zeta })P_{0}.  \label{VP}
\end{equation}%
\qquad The G$^{2}$CEq contains the general $P$, while the special case using 
$P_{0\ }$yields the GCEq.

For the most general situation, the G$^{2}$CEq can be written as a
differential equation for the function $u=G(\zeta ,\overline{\zeta })$:%
\begin{equation}
\overline{\eth }^{2}G\equiv \partial _{_{\overline{\zeta }%
}}(V^{2}P_{0}^{2}\partial _{\overline{\zeta }}G)=\sigma (G,\zeta ,\overline{%
\zeta }),  \label{G^2CEq.I}
\end{equation}%
or%
\begin{equation}
\ \ P_{0}^{2}\partial _{\overline{\zeta }}^{2}G+2[P_{0}^{2}V^{-1}\partial
_{_{\overline{\zeta }}}V+P_{0}\zeta ]\partial _{\overline{\zeta }%
}G=V^{-2}\sigma (G,\zeta ,\overline{\zeta }).  \label{G^2CEq}
\end{equation}%
When $V=1\ $we have the GCEq: 
\begin{equation}
\overline{\eth }_{0}^{2}G\equiv \partial _{_{\overline{\zeta }%
}}(P_{0}^{2}\partial _{\overline{\zeta }}G)=\sigma (G,\zeta ,\overline{\zeta 
}).  \label{GCEq}
\end{equation}%
Further, when the arbitrary spin-weight 2 function $\sigma (G,\zeta ,\bar{%
\zeta})\ $vanishes, we have the homogeneous GCEq:%
\begin{equation}
\partial _{_{\overline{\zeta }}}(P_{0}^{2}\partial _{\overline{\zeta }}G)=0.
\label{hGCEq}
\end{equation}

In the following section it will be shown that the G$^{2}$CEq can, by a
(non-obvious) coordinate transformation, be transformed into Eq.(\ref{GCEq}%
). We can simply describe the solutions (and its properties) of Eq.(\ref%
{GCEq}): it then follows that they also are true for Eq.(\ref{G^2CEq}). \
Chief among these properties is the fact the solution space to (\ref{GCEq})
is a complex 4-manifold with a natural Einstein metric; hence, taking any
curve in this solution space produces a 1-parameter family of good cut
functions which in turn describe a foliation of 3-surface $\mathfrak{N}$.

\begin{remark}
Solutions to the GCEq or G$^{2}$CEq, $u=G(\zeta ,\bar{\zeta})$, known as
\textquotedblleft good cut functions,\textquotedblright\ describe
cross-sections of $\mathfrak{N\ }$that are referred to as "good cuts". From
the tangent tangents to these good cuts, $L=\overline{\eth }G,\ $one can
construct null directions, (out of $\mathfrak{N}$), into the space-time
iteself\ that determine a NGC whose shear vanishes at $\mathfrak{N.\ }$(See
the following section.) When $\mathfrak{N=I}^{+}$, these are asymptotically
shear-free NGCs; when $\mathfrak{N}$ is a non-expanding horizon, these are
vacuum-horizon-shear-free NGCs.
\end{remark}

\ \ 

\subsection{Solution Space of the GCEq}

In this and the following subsection we will be concerned \textit{only} with
the GCEq\ (\ref{GCEq}) and its solutions.

The key fact about the solution space to the GCEq is that it forms a complex
4-manifold, known as $\mathcal{H}$-space, for sufficiently regular $\sigma
(G,\zeta ,\bar{\zeta})$ (which is assumed here). \ This manifold of
solutions can be shown to possess a vacuum Einstein metric with
anti-self-dual Weyl tensor. \ Although the rigorous proof of the existence
and properties of $\mathcal{H}$-space requires the use of Kodaira
deformation theory and Penrose's non-linear graviton construction \cite%
{HNPT78}, we can give a simple intuitive argument here.\ The solutions will
be written as%
\begin{equation}
u=G(z^{a},\zeta ,\overline{\zeta }),  \label{solution}
\end{equation}%
with $z^{a}\ $(appearing as four constants of integration) the $\mathcal{H}$%
-space coordinates.

Since, from the properties of the (sphere) $\eth _{0}$-operator, one
immediately has that the general \textit{regular} solution of the \textit{%
homogeneous} equation (\ref{hGCEq}) is given by the four parameter function%
\begin{eqnarray}
G_{0}(z^{a},\zeta ,\overline{\zeta }) &=&z^{a}l_{a}(\zeta ,\overline{\zeta }%
)=\frac{\sqrt{2}}{2}z^{0}+\frac{1}{2}z^{i}Y_{1i}^{0}(\zeta ,\overline{\zeta }%
),  \label{homog.sol} \\
l_{a}(\zeta ,\overline{\zeta }) &=&\frac{\sqrt{2}}{2P_{0}}(1+\zeta \overline{%
\zeta },\ \zeta +\bar{\zeta},\ i\bar{\zeta}-i\zeta ,\ -1+\zeta \bar{\zeta}).
\nonumber
\end{eqnarray}%
The inhomogeneous equation can then be rewritten as the integral equation%
\begin{equation}
G=G_{0}(z^{a},\zeta ,\overline{\zeta })+\int \sigma ^{0}(G,\eta ,\bar{\eta}%
)K_{0,-2}^{+}(\eta ,\bar{\eta}\text{:}\zeta ,\overline{\zeta })dS_{\eta },
\label{iteration}
\end{equation}%
with 
\begin{eqnarray}
K_{0,-2}^{+}(\zeta ,\overline{\zeta },\eta ,\bar{\eta}) &\equiv &-\frac{1}{%
4\pi }\frac{(1+\overline{\zeta }\eta )^{2}(\eta -\zeta )}{(1+\zeta \overline{%
\zeta })(1+\eta \bar{\eta})(\bar{\eta}-\overline{\zeta })},  \label{kernel}
\\
dS_{\eta } &=&4i\frac{d\eta \wedge d\bar{\eta}}{(1+\eta \bar{\eta})^{2}},
\label{area}
\end{eqnarray}%
where $K_{0,-2}^{+}(\zeta ,\tilde{\zeta},\eta ,\tilde{\eta})$ is the Green's
function for the $\eth _{0}^{2}$-operator \cite{IKN89}. By iterating this
equation, with $G_{0}(z^{a},\zeta ,\tilde{\zeta})=z^{a}l_{a}(\zeta ,%
\overline{\zeta })$ being the zeroth iterate,

\begin{equation}
G_{n}(\zeta ,\overline{\zeta })=z^{a}l_{a}(\zeta ,\overline{\zeta }%
)+\int_{S^{2}}K_{0,-2}^{+}(\zeta ,\overline{\zeta },\eta ,\bar{\eta})\sigma
(G_{n-1},\eta ,\bar{\eta})dS_{\eta },  \label{iteration2}
\end{equation}%
one easily sees how the four $z^{a}$ enter the solution: the four constants
originate from the solution to the homogeneous equation.

We thus have the result that the solutions to the GCEq. 
\begin{equation}
u=G(z^{a},\zeta ,\overline{\zeta }),  \label{regular}
\end{equation}%
defines a four-parameter family of cuts of $\mathfrak{N}$, each cut labeled
by the $\mathcal{H}$-space points, $z^{a}.$ By choosing an arbitrary
analytic curve in the $\mathcal{H}$-space, $z^{a}=\xi ^{a}(\tau ),\ $with $%
\tau \ $an arbitrary complex parameter, we have a one-parameter family of
cuts of $\mathfrak{N}$

\[
u=Z(\tau ,\zeta ,\overline{\zeta })=G(\xi ^{a}(\tau ),\zeta ,\overline{\zeta 
}). 
\]%
\qquad Each choice of the curve $z^{a}=\xi ^{a}(\tau )\ $yields an
asympotically shear-free NGC by the following construction:

Considering $\mathfrak{N\ }$as embedded in a Lorentzian space-time, at each
point of $\mathfrak{N\ }$we can construct its past (or future) light-cone.
The sphere of null directions can be coordinitized by complex stereographic
coordinates, ($L,\overline{L}$). \ A field of null directions pointing
backward (or forward) from $\mathfrak{N\ }$can be written as%
\begin{equation}
L=L(\mathfrak{N})=L(u,\zeta ,\overline{\zeta }).  \label{L field}
\end{equation}%
\qquad It is known that the field of null directions that is given
parametrically on $\mathfrak{N\ }$by%
\begin{eqnarray*}
L(u,\zeta ,\overline{\zeta }) &=&\eth _{0}Z(\tau ,\zeta ,\overline{\zeta }),
\\
u &=&Z(\tau ,\zeta ,\overline{\zeta }),
\end{eqnarray*}%
describes the null direction field of an asymptotically shearfree NGC \cite%
{AN72,Newman76,KNSS07,AKN09}. When we are dealing with the homogeneous
equation on $\mathfrak{I}$ in Minkowski space, the NGC turns out to be
shear-free everywhere.

We point out that the solution space comes naturally with the ($\mathcal{H}$%
-space) complex metric \cite{Newman76}%
\begin{eqnarray}
ds_{(\mathcal{H})}^{2} &=&g_{(\mathcal{H})ab}\,dz^{a}dz^{b}\equiv \left( 
\frac{1}{8\pi }\int_{S^{2}}\frac{dS}{(dG)^{2}}\right) ^{-1},  \label{metric}
\\
dG &=&\nabla _{a}G\ dz^{a}  \nonumber \\
dS &=&4i\frac{d\zeta {\small \wedge }d\overline{\zeta }}{(1+\zeta \overline{%
\zeta })^{2}},  \nonumber
\end{eqnarray}%
that is Ricci flat and has anti-self-dual conformal (Weyl) curvature. \ For
the special case of solutions to the homogeneous GCEq, this metric reduces
to the complex Minkowski metric.

\qquad

\subsection{The Good Cut Equation and Twistors}

The study of the GCEq is intimately related to Penrose's twistor theory. \
Although we will not provide an extensive review of twistor theory here, we
will include the briefest of overviews to set the stage for the following
discussion. \ The interested reader need only consult \cite{HT85,PR86} for a
more in-depth introduction and discussion. \ For our purposes, twistor space
is the complex projective 3-space $\mathbb{PT\simeq CP}^{3}$, charted with
homogeneous coordinates $Z^{\alpha }=(\omega ^{A},\pi _{A^{\prime }})$,
where $\omega $ and $\pi $ are un-primed and primed Weyl spinors
respectively. \ A (projective) twistor is any point $Z^{\alpha }\in \mathbb{%
PT}$. \ Twistor space is related to points $x$ in complex Minkowski
space-time by the incidence relation:%
\begin{equation}
\omega ^{A}=ix^{AA^{\prime }}\pi _{A^{\prime }},  \label{inc}
\end{equation}%
where $x^{AA^{\prime }}$ is the usual spinor representation (where a vector
index is replaced by a pair of primed and un-primed spinor indicies),%
\[
x^{AA^{\prime }}=\frac{1}{\sqrt{2}}\left( 
\begin{array}{cc}
t+x & y-iz \\ 
y+iz & t-x%
\end{array}%
\right) .
\]%
Eq.(\ref{inc}) can be used to determine that a point in $\mathbb{PT}$
corresponds to a null geodesic in complex Minkowski space, while a point $x$
in complex Minkowski space corresponds to a line $L_{x}\simeq \mathbb{CP}%
^{1}\subset \mathbb{PT}$. \ The geometry of real Minkowski space-time is
recovered on null twistor space, defined as%
\[
\mathbb{PN}=\{Z^{\alpha }\in \mathbb{PT}:\omega ^{A}\bar{\pi}_{A}+\pi
^{A^{\prime }}\bar{\omega}_{A^{\prime }}=0\}.
\]%
In other words, $Z^{\alpha }$ corresponds to a real null geodesic in
Minkowski space-time if and only if $Z^{\alpha }\in \mathbb{PN}$; and $L_{x}$
corresponds to a real point in Minkowski space-time if and only if $%
L_{x}\subset \mathbb{PN}$.

We can also chart $\mathbb{PT}$ with non-homogeneous coordinates, which are
most useful when studying the GCEq.\ Assuming that $\pi _{A^{\prime }}\neq 0$
(which corresponds to excluding the point at infinity by (\ref{inc})) and
working on a patch where $\pi _{0^{\prime }}\neq 0$, we can write \cite%
{KNP77}:%
\[
(\omega ^{A},\pi _{A^{\prime }})=(ir\mu ^{0},-ir\mu ^{1},r,-r\zeta ), 
\]%
where $r\in \mathbb{C}$ is a common scaling factor and $\zeta $ can be shown
to be equivalent to the complex stereographic angle on $S^{2}$ introduced
earlier. \ This means that $(\zeta ,\mu ^{0},\mu ^{1})$ can be interpreted
as coordinates on $\mathbb{PT\cong CP}^{3}$ via%
\begin{eqnarray*}
\zeta &=&-\frac{\pi _{1^{\prime }}}{\pi _{0^{\prime }}}, \\
\mu ^{0} &=&-i\frac{\omega ^{0}}{\pi _{0^{\prime }}}, \\
\mu ^{1} &=&i\frac{\omega ^{1}}{\pi _{0^{\prime }}}.
\end{eqnarray*}

A curved twistor space $\mathbb{P}\mathcal{T}$ can also be constructed for
any complex space-time that is Ricci flat with anti-self-dual conformal
curvature by the non-linear graviton construction (c.f., \cite{HT85}). \
Such curved twistor spaces have a similar correspondence with null geodesics
and points in the complex space-time, although the curves $L_{x}$ will no
longer be lines, as in (\ref{inc}).

We now discuss the relationship between the above discussion of twistor
space and the GCEq.

Starting with the homogeneous GCEq, twistor space can be constructed in the
following manner. Treating the variable $\zeta =\zeta _{0}\ $for the moment
as a\ fixed constant, the homogeneous GCEq (\ref{hGCEq}) becomes a
second-order ODE for $u=G(\overline{\zeta }).\ $Its solution is determined$\ 
$by two initial conditions: the value of $G\ $and its first derivative at $%
\overline{\zeta }\ $equal to the complex$\ $conjugate of $\zeta \ $(denoted
by $\overline{\zeta }_{0}$)$,\ $i.e., at $u_{0}=G(\overline{\zeta }_{0})\ $%
and $L_{0}=(1+\overline{\zeta }_{0}\zeta _{0})\partial _{\overline{\zeta }%
_{0}}G(\overline{\zeta }_{0}).\ $The curves so-determined are defined as a
projective twistors, with $\mathbb{PT}$ being the collection of all such
curves. We then adopt ($\zeta _{0},u_{0},L_{0}$) as local coordinates on $%
\mathbb{PT}$, with the relationship to the standard twistor coordinates
given earlier by%
\[
(\zeta _{0},\mu _{0},\mu _{1})=(\zeta _{0},\ u_{0}-\overline{\zeta }%
_{0}L_{0},\ \zeta _{0}u_{0}+L_{0}). 
\]%
These relations come directly from the integration of the ODE:%
\begin{eqnarray*}
\partial _{\overline{\zeta }}(1+\zeta _{0}\overline{\zeta })^{2}\partial _{%
\overline{\zeta }}G &=&0,\ \  \\
\partial _{\overline{\zeta }}G &=&(1+\zeta _{0}\overline{\zeta })^{-2}\alpha
_{0}, \\
\overline{L} &=&\overline{\eth }G\equiv (1+\zeta _{0}\overline{\zeta }%
)\partial _{\overline{\zeta }}G=(1+\zeta _{0}\overline{\zeta })^{-1}\alpha
_{0}, \\
&\Longrightarrow &u=G=\alpha _{1}-\zeta _{0}^{-1}(1+\zeta _{0}\overline{%
\zeta })^{-1}\alpha _{0}.
\end{eqnarray*}%
The pair of integration constants, ($\alpha _{0},\alpha _{1}$),\ are
determined directly in terms of the initial conditions ($u_{0},L_{0}$).
Defining $(\mu _{0},\mu _{1})\ $by 
\[
\alpha _{0}=\mu _{1}-\zeta \mu _{0},\ \ \alpha _{1}=\mu _{1}\zeta ^{-1}, 
\]%
we have 
\begin{eqnarray*}
u &=&(\mu _{1}\overline{\zeta }+\mu _{0})(1+\zeta _{0}\overline{\zeta })^{-1}
\\
\overline{L} &=&(\mu _{1}-\zeta _{0}\mu _{0})(1+\zeta _{0}\overline{\zeta }%
)^{-1} \\
u_{0} &=&(\mu _{1}\overline{\zeta }_{0}+\mu _{0})(1+\zeta _{0}\overline{%
\zeta }_{0})^{-1} \\
\overline{L}_{0} &=&(\mu _{1}-\zeta _{0}\mu _{0})(1+\zeta _{0}\overline{%
\zeta }_{0})^{-1}.
\end{eqnarray*}%
\qquad

Suppose two different twistors were chosen with the local coordinates ($%
\zeta _{0},u_{0},L_{0}$)\ and ($\zeta _{1},u_{1},L_{1}$), and their
respective solution curves $G_{1,2}$ equated with an arbitrary four
parameter regular solution of the form (\ref{homog.sol}) 
\[
G_{0}(z^{a},\zeta ,\overline{\zeta })=z^{a}l_{a}(\zeta ,\overline{\zeta })=%
\frac{\sqrt{2}}{2}z^{0}+\frac{1}{2}z^{i}Y_{1i}^{0}(\zeta ,\overline{\zeta }),
\]%
at $\zeta =\zeta _{0}\ $and $\zeta =\zeta _{1}$.$\ $This yields four linear
algebraic equations to determine the four$\ $coordinates $z^{a}\ $in terms
of the four ($u_{0},L_{0},u_{1},L_{1}$). \ This construction is totally
equivalent to the use of the twistor incidence relationship (\ref{inc}) to
determine the space-time points $x^{AA^{\prime }}\ $from a pair of
projective twistors, since the two twistors ($\zeta _{0},u_{0},L_{0}$)\ and (%
$\zeta _{1},u_{1},L_{1}$) uniquely determine a line $L_{x}\subset \mathbb{PT}
$. \ This is a linear relationship, in that the choice of any pair of
twistors determines both the space-time point $x^{AA^{\prime }}\ $and a line
in projective twistor space , where any pair of points on the line determine
the same point $x^{AA^{\prime }}$.

The attempt to apply this construction to the inhomogeneous equation, Eq.(%
\ref{GCEq}) fails; the relationship is no longer linear. \ One obtains
instead a differential equation describing a (non-linear) curve in a curved
twistor space such that any point\ and its tangent vector on the curve
determines a point in $\mathcal{H}$-space. The argument we give for this is
very heuristic in the sense we are assuming (without proof) that several
implicit algebraic equations can be inverted. When the situation is
sufficiently close to that of the homogeneous equation (i.e., for suitably
small $\sigma (u,\zeta ,\bar{\zeta})$), this should not be a problem.

By rewriting Eq.(\ref{GCEq}) as%
\[
P_{0}^{2}\partial _{_{\overline{\zeta }}}\partial _{\overline{\zeta }%
}G+2P_{0}\zeta \partial _{\overline{\zeta }}G=\sigma (G,\zeta ,\overline{%
\zeta }), 
\]%
or in the compressed form%
\begin{eqnarray}
\partial _{\overline{\zeta }}^{2}G &=&S(G,\partial _{\overline{\zeta }%
}G,\zeta ,\overline{\zeta }),  \label{GCEq2} \\
S &\equiv &P_{0}^{-2}\sigma (G,\zeta ,\overline{\zeta })-2P_{0}^{-1}\zeta
\partial _{\overline{\zeta }}G,  \nonumber
\end{eqnarray}%
we see that there are two different types of solutions:

\qquad The first comes from the condition that $\zeta =\zeta _{0}\ $is taken
as a constant and Eq.(\ref{GCEq2}) can be treated as a second order ODE for $%
G$ as a function of $\overline{\zeta }\ $whose solution depends on two
constants of integration, ($\alpha ,\beta $). They can be taken as the
initial value and first derivative of $G\ $at some arbitrary point $%
\overline{\zeta }=\widetilde{\zeta }_{0}.$ On a fiducial (or special) curve
we take $\widetilde{\zeta }_{0}\ $to be the complex conjugate of $\zeta
_{0}:\ $ $\widetilde{\zeta }_{0}=\overline{\zeta }_{0}.\ $The solution will
then be written as 
\begin{equation}
u=G^{(1)}(\alpha (\overline{\zeta }_{0}),\beta (\overline{\zeta }_{0}),\zeta
_{0},\overline{\zeta }).  \label{G^1}
\end{equation}

This curve in ($u,\overline{\zeta }$)\ space, labeled by $(\alpha ,\beta
,\zeta _{0}),$ is identified as a twistor with local coordinates ($\alpha (%
\overline{\zeta }_{0}),\beta (\overline{\zeta }_{0}),\zeta _{0}$)$.\ $(It
should be noted that if the initial value point, $\overline{\zeta }_{0},$ on
the twistor curve is changed, the new values of $\alpha \ $and $\beta \ $are
easily found.) By freeing up $\zeta _{0},\ $i.e., allowing $\zeta
_{0}\rightarrow $ $\zeta ,\ $to vary and letting $\alpha ,\beta \ $be
functions of both $\zeta $ and the fixed initial value point$\ \overline{%
\zeta }_{0},\ $we obtain a one-parameter family of twistors (or a
twistor-space curve), 
\begin{equation}
u=G^{(1)}(\alpha (\zeta ,\overline{\zeta }_{0}),\beta (\zeta ,\overline{%
\zeta }_{0}),\zeta ,\overline{\zeta }).
\end{equation}

The second type of solutions are the \textit{regular }ones, eq.(\ref{regular}%
), that depend on four constants, the $\mathcal{H}$-space coordinates\textit{%
, }$z^{a}:$\textit{\ }%
\begin{equation}
u=G^{(2)}(z^{a},\zeta ,\overline{\zeta }).  \label{regular2}
\end{equation}%
Holding the $z^{a}\ $fixed and varying $\zeta $, we obtain a one-parameter
family of twistors (i.e. a curve in $\mathbb{P}\mathcal{T}$).

The question is: how can the two functions $\alpha (\zeta ,\overline{\zeta }%
_{0}),\ \beta (\zeta ,\overline{\zeta }_{0})\ $be chosen so that two sets of
solutions coincide? \ 

This is accomplished by first equating the two types of solution and their
first $\overline{\zeta }\ $derivatives at $\overline{\zeta }=\overline{\zeta 
}_{0}$:

\begin{eqnarray}
G^{(1)}(\alpha (\zeta ,\overline{\zeta }_{0}),\beta (\zeta ,\overline{\zeta }%
_{0}),\zeta ,\overline{\zeta }_{0}) &=&G^{(2)}(z^{a},\zeta ,\overline{\zeta }%
_{0}),  \label{A} \\
\partial _{\overline{\zeta }}G^{(1)}(\alpha (\zeta ,\overline{\zeta }%
_{0}),\beta (\zeta ,\overline{\zeta }_{0}),\zeta ,\overline{\zeta }_{0})
&=&\partial _{\overline{\zeta }}G^{(2)}(z^{a},\zeta ,\overline{\zeta }_{0}).
\label{B}
\end{eqnarray}%
We thus have a pair of implicit equations whose algebraic solution for $%
\alpha \ $and $\beta \ $has the form:%
\begin{eqnarray}
\alpha (\zeta ,\overline{\zeta }_{0}) &=&A(z^{a},\zeta ,\overline{\zeta }%
_{0}),  \label{A*} \\
\beta (\zeta ,\overline{\zeta }_{0}) &=&B(z^{a},\zeta ,\overline{\zeta }%
_{0}).  \label{B*}
\end{eqnarray}

It is easy to derive, by the following argument, a (parametric) pair of
second order ODEs for $\alpha $ and $\beta $ where (\ref{A*}-\ref{B*}) are
the solutions and $z^{a}$ are the constants of integration. \ First take the 
$\zeta $ derivative of (\ref{A*}) and (\ref{B*}):%
\begin{eqnarray}
\alpha ^{\prime }(\zeta ,\overline{\zeta }_{0}) &=&\partial _{\zeta
}A(z^{a},\zeta ,\overline{\zeta }_{0}),  \label{A**} \\
\beta ^{\prime }(\zeta ,\overline{\zeta }_{0}) &=&\partial _{\zeta
}B(z^{a},\zeta ,\overline{\zeta }_{0}).  \label{B**}
\end{eqnarray}%
The two pairs, Eqs.(\ref{A*}-\ref{B*}) and (\ref{A**}-\ref{B**}), determine
implicitly%
\begin{equation}
z^{a}=\mathcal{Z}^{a}(\alpha ,\beta ,\alpha ^{\prime },\beta ^{\prime
},\zeta ,\overline{\zeta }_{0}).  \label{za}
\end{equation}

Finally taking the $\zeta \ $derivative of (\ref{A**}) and (\ref{B**}):%
\begin{eqnarray}
\alpha ^{\prime \prime }(\zeta ,\overline{\zeta }_{0}) &=&\partial _{\zeta
}\partial _{\zeta }A(z^{a},\zeta ,\overline{\zeta }_{0}), \\
\beta ^{\prime \prime }(\zeta ,\overline{\zeta }_{0}) &=&\partial _{\zeta
}\partial _{\zeta }B(z^{a},\zeta ,\overline{\zeta }_{0}),
\end{eqnarray}%
and eliminating $z^{a},\ $via (\ref{za}), leaves the pair%
\begin{eqnarray}
\alpha ^{\prime \prime }(\zeta ,\overline{\zeta }_{0}) &=&\mathcal{A}(\alpha
,\beta ,\alpha ^{\prime },\beta ^{\prime },\zeta ,\overline{\zeta }_{0}),
\label{alph''} \\
\beta ^{\prime \prime }(\zeta ,\overline{\zeta }_{0}) &=&\mathcal{B}(\alpha
,\beta ,\alpha ^{\prime },\beta ^{\prime },\zeta ,\overline{\zeta }_{0}).
\label{beta''}
\end{eqnarray}%
The solutions determine, for each set of constants, $z^{a},\ $a curve in the
asymptotic twistor space. \ 

Though it might be (and probably should be) possible to construct the $%
\mathcal{A\ }$and $\mathcal{B}$\ directly from the$\ $GCEq. (i.e., from $%
\sigma (u,\zeta ,\overline{\zeta })$), we do not at the present know how to
do this. Nevertheless it is nice to see that such curves do exist and
determine points in the $\mathcal{H}$-space. \ Hence, we have confirmed
precisely what is known from twistor theory: a point in $\mathcal{H}$-space
(a vacuum, anti-self-dual complex space-time) corresponds to a curve in a
curved twistor space. \ Indeed, since $\mathcal{H}$-space reduces to complex
Minkowski space in the case of the homogeneous GCEq, we saw (as expected)
that points in this trivial $\mathcal{H}$-space corresponded to lines in $%
\mathbb{PT}$.

\section{Equivalence of Good Cut Equations: From the G$^{2}$CEq to the GCEq}

In this section we show how, by a coordinate transformation of the
(independent) complex stereographic coordinates ($\zeta ,\overline{\zeta }$%
), the generalized GCEq (G$^{2}$CEq) can be transformed into the GCEq. It
must be remembered from our notation that $\overline{\zeta }^{\ast }($or$\ 
\overline{\zeta })\ $is close to, but is not necessarily, the complex
conjugate of $\zeta ^{\ast }\ ($or $\zeta )$.

We first rewrite the GCEq with stereographic coordinates ($\zeta ^{\ast },%
\overline{\zeta }^{\ast }$) as%
\begin{eqnarray}
\overline{\eth }_{0\ast }^{2}G &=&\partial _{_{\overline{\zeta }^{\ast
}}}(P_{0}^{\ast 2}\partial _{_{\overline{\zeta }^{\ast }}}G)=\sigma ^{\ast
}(G,\zeta ^{\ast },\overline{\zeta }^{\ast }),  \label{C} \\
P_{0}^{\ast } &=&1+\zeta ^{\ast }\overline{\zeta }^{\ast },  \label{P*}
\end{eqnarray}%
and the G$^{2}$CEq as%
\begin{equation}
\overline{\eth }^{2}G=\partial _{_{\overline{\zeta }}}(V^{2}P_{0}^{2}%
\partial _{\overline{\zeta }}G)=\sigma (G,\zeta ,\overline{\zeta }).
\label{D}
\end{equation}

We now apply the coordinate transformation

\begin{eqnarray}
\ \ \overline{\zeta }^{\ast } &=&\frac{\overline{\zeta }+W}{1-W\zeta }\equiv
N(\zeta ,\overline{\zeta }),\ \   \label{CT} \\
\zeta ^{\ast } &=&\zeta ,\ \ 
\end{eqnarray}%
with $W$ (a spin-weight 1 function) defined from%
\begin{eqnarray*}
V^{-2} &=&1+\overline{\eth }_{0}W=1+P_{0}\partial _{_{\overline{\zeta }%
}}W-W\zeta , \\
P_{0} &=&1+\zeta \overline{\zeta },
\end{eqnarray*}%
to Eq.(\ref{D}). \ Substituting the derived relations,%
\begin{eqnarray*}
P_{0}^{\ast } &=&1+\zeta \overline{\zeta }^{\ast }=\frac{1+\zeta \overline{%
\zeta }}{1-W\zeta }=\frac{P_{0}}{1-W\zeta }, \\
\partial _{_{\overline{\zeta }}}G &=&\partial _{_{\overline{\zeta }^{\ast
}}}G\cdot \partial _{_{\overline{\zeta }}}N, \\
\partial _{_{\overline{\zeta }}}^{2}G &=&\partial _{_{\overline{\zeta }%
^{\ast }}}^{2}G\cdot (\partial _{_{\overline{\zeta }}}N)^{2}+\partial _{_{%
\overline{\zeta }^{\ast }}}G\cdot \partial _{_{\overline{\zeta }}}^{2}N, \\
\partial _{_{\overline{\zeta }}}N &=&\frac{V^{-2}-W\zeta }{(1-W\zeta )^{2}},
\\
\partial _{_{\overline{\zeta }}}^{2}N &=&\frac{2\zeta \lbrack
V^{-2}-1]\partial _{_{\overline{\zeta }}}W}{(1-W\zeta )^{3}}+\frac{\zeta
\partial _{_{\overline{\zeta }}}W}{(1-W\zeta )^{2}}-\frac{2V^{-3}\partial
_{_{\overline{\zeta }}}V}{(1-W\zeta )^{2}},
\end{eqnarray*}%
into Eq.(\ref{D}), we have, after a bit of algebra,

\[
\overline{\eth }_{0\ast }^{2}G=\partial _{_{\overline{\zeta }^{\ast
}}}(P_{0}^{\ast 2}\partial _{_{\overline{\zeta }^{\ast }}}G)=F(\zeta ^{\ast
},\overline{\zeta }^{\ast })\sigma (G,\zeta (\zeta ^{\ast },\overline{\zeta }%
^{\ast }),\overline{\zeta }^{\ast }((\zeta ^{\ast },\overline{\zeta }^{\ast
}))\equiv \sigma ^{\ast }(G,\zeta ^{\ast },\overline{\zeta }^{\ast }), 
\]%
namely Eq.(\ref{C}), the GCEq.

Hence, we see that the G$^{2}$CEq is really equivalent to the GCEq via the
coordinate transformation (\ref{CT}). \ This means that the study of the G$%
^{2}$CEq on a general 3-surface $\mathfrak{N}$ can be reduced to the study
of the properties of the GCEq on a 3-surface whose cross-sections are metric
spheres. \ Although the form of the coordinate transformation is far from
obvious, this equivalence is not totally un-expected, since the good cut
equation is known to be conformally invariant.

\subsection{Application: Vacuum Non-Expanding Horizons}

\bigskip In \cite{AN09}, it was shown that the condition for a neighborhood
of a vacuum non-expanding horizon $\mathfrak{H}$\footnote{%
A vacuum non-expanding horizon is a null 3-submanifold in a Lorentzian
space-time which has vanishing divergence and shear, is topologically $%
\,S^{2}\times \mathbb{R}$, and is fibered over $S^{2}$ by null curves with
the vacuum Einstein equations holding in a neighborhood of the horizon.} to
be foliated by a null geodesic congruence whose shear vanishes at the
horizon takes the form of a time-independent G$^{2}$CEq:%
\[
\eth ^{2}G=\sigma (\zeta ,\bar{\zeta}),
\]%
where the right-hand side does not depend on $u$. \ It isn't hard to prove
that the solution space to this equation is a complex 4-manifold, but it is
unclear if this solution space posesses the ordinary $\mathcal{H}$-space
metric. \ Equivalently, we want to know: is there a twistor space associated
with the generalized good cut equation on the horizon $\mathfrak{H}$?

Answering this question is now almost trivial in light of our previous
discussion. \ Since we continue to work with the complexified surface, a
complex supertranslation%
\[
u\rightarrow u+f(\zeta ,\bar{\zeta}) 
\]%
shifts the function $\sigma $ by Sachs' Theorem:%
\[
\sigma (\zeta ,\bar{\zeta})\rightarrow \sigma (\zeta ,\bar{\zeta})+\eth
^{2}f(\zeta ,\bar{\zeta}). 
\]%
Hence, we can choose $f$ to set $\sigma =0$ on the entire horizon. \ This
leaves us with the homogeneous G$^{2}$CEq:%
\[
\eth ^{2}G=0. 
\]%
But by the coordinate transformation (\ref{CT}), we know that this is
equivalent to the homogeneous GCEq, which we saw had complex Minkowski space
as its solution space. \ From Section 2.2, we know that complex Minkowski
space corresponds to (flat) twistor space $\mathbb{PT}$, so it follows that
there is a flat twistor space associated with horizon-shear-free NGCs
intersecting any vacuum non-expanding horizon.

\section{Conclusion}

In this note we have studied an old topic: the good cut equation and its
solution space. \ In Section 2, we saw how both flat and curved twistor
space twistor could be used to explicitly understand the relationship
between points in $\mathcal{H}$-space and solutions to the good cut equation
on a 3-surface $\mathfrak{N}$. \ In the setting where $\mathfrak{N}$ has
non-metric sphere cross-sections (such as non-expanding horizons embedded in
space-time), generalized good cut equations arise, which differ from the
ordinary GCEq in the definition of the $\eth $-operator on the $S^{2}$
portion of the topology. \ Section 3 demonstrated that these generalized
good cut equations are related to the GCEq by a simple coordinate
transformation on the complex stereographic coordinates $(\zeta ,\bar{\zeta})
$. \ This means that the study of the G$^{2}$CEq on $\mathfrak{N}$ with
arbitrary conformal factor on the sphere topology reduces to the study of
the normal GCEq on metric 2-spheres. \ As an example, we saw that this
observation immediately implies that the solution space associated to the G$%
^{2}$CEq on a vacuum non-expanding horizon is complex Minkowski space, and
that the corresponding twistor space is the flat $\mathbb{PT}$. \ 

In particular, this indicates that recently developed physical
identification theories based on the study of solutions to the GCEq on $%
\mathfrak{I}^{+}$ could be adapted to non-conformal local 3-surfaces
embedded in space-time (c.f., \cite{AKN09}). \ In the future, we hope to
apply these results to prior studies of non-expanding horizons (e.g., \cite%
{AN09}) in the hope of developing a local physical identification theory
which could identify physical quantities such as mass, linear momentum, and
angular momentum flux at null 3-surfaces in space-time.\ 

\bibliography{GGCEq}
\bibliographystyle{acm}

\end{document}